\documentclass[aps,pra,twocolumn,showpacs]{revtex4}

\usepackage{graphicx}
\newcommand{\tr}{\mbox{Tr} }
\newcommand{\ket}[1]{\left | #1 \right \rangle}
\newcommand{\bra}[1]{\left \langle #1 \right |}
\newcommand{\proj}[1]{\ket{#1}\!\!\bra{#1}}

\begin{document}
\title{Entanglement fidelity and measurement of entanglement preserving in
quantum processes}
\author{Yang Xiang}
\email{njuxy@sina.com}
\author{Shi-Jie Xiong }

\affiliation{National Laboratory of Solid State Microstructures and
Department of Physics, Nanjing University, Nanjing 210093, China}
\date{\today}
\begin{abstract}
The entanglement fidelity provides a measure of how well the
entanglement between two subsystems is preserved in a quantum
process. By using a simple model we show that in some cases this
quantity in its original definition fails in the measurement of the
entanglement preserving. On the contrary, the modified entanglement
fidelity, obtained by using a proper local unitary transformation on
a subsystem, is shown to exhibit the behavior similar to that of the
concurrence in the quantum evolution.
\end{abstract}

\maketitle




Quantum entanglement is a key element for applications of quantum
communications and quantum information. A complete discussion of
this has been given in Ref. \cite{horo1}. Characterizing and
quantifying the entanglement is a fundamental issue in quantum
information theory. For pure and mixed states of two qubits this
problem about the description of the entanglement has been well
elucidated
\cite{werner,bennett1,bennett2,wootters1,uhlmann,wootters2}.
Recently, Jordan \textsl{et al.} \cite{jordan} considered two
entangled qubits, one of which interacts with a third qubit named as
a control one that is never entangled with either of the two
entangled qubits. They found that the entanglement of these two
qubits can be both increased and decreased by the interaction with
the control qubit on just one of them. If we regard the control
qubit as an environment and the state of the qubit interacting with
the control qubit as the information source, this example is just a
model for the time evolution of quantum information via a noisy
quantum channel originating from the interaction with the control
qubit. Schumacher \cite{schumacher} and Barnum \textsl{et al.}
\cite{barnum} have investigated a general situation where $R$ and
$Q$ are two quantum systems and the joint system $RQ$ is initially
prepared in a pure entangled state $|\Psi^{RQ}\rangle$. The system
$R$ is dynamically isolated and has a zero internal Hamiltonian,
while the system $Q$ undergoes some evolution that possibly involves
interaction with the environment. The evolution of $Q$ might
represent a transmission process via some quantum channel for the
quantum information in $Q$. They introduced a fidelity
$F_{e}=\langle\Psi^{RQ}|\rho^{RQ'}|\Psi^{RQ}\rangle$, which is the
probability that the final state $\rho^{RQ'}$ would pass a test
checking whether it agrees with the initial state
$|\Psi^{RQ}\rangle$. This quantity is called as entanglement
fidelity (referred hereafter as EF). The EF can be defined entirely
in terms of the initial state $\rho^{Q}$ and the evolution of system
$Q$, so EF is related to a process, specified by a quantum operation
$\varepsilon^{Q}$, which we shall discuss later in more details,
acting on some initial state $\rho^{Q}$. Thus, the EF can be denoted
by a function form $F_{e}(\rho^{Q},\varepsilon^{Q})$. The EF is
usually used to measure how well the state $\rho^{Q}$ is preserved
by the operation $\varepsilon^{Q}$ and to identify how well the
entanglement of $\rho^{Q}$ with other systems is preserved by the
operation of $\varepsilon^{Q}$. The complete discussion of EF can be
seen in \cite{nielsen,schumacher}. In the present work we will
investigate the following question: Is EF a good measurement of the
entanglement preserving? Using the example of Jordan {\it et al.},
we find that in some cases EF defined above completely fails for
measuring the entanglement preserving though it may be a good
measurement of the entanglement preserving in the case of slight
noise. We also find that in order to make the EF indeed equivalent
to an entanglement measure the modified entanglement fidelity (MEF)
should be used. Some detailed discussions about the MEF have been
given in \cite{schumacher,nielsen2,kret}. Recently, Surmacz {\it et
al.} \cite{karl2} have investigated the evolution of the
entanglement in a quantum memory and showed that the MEF can be used
to measure how well a quantum memory setup can preserve the
entanglement between a qubit undergoing the memory process and an
auxiliary qubit. For the example of Jordan {\it et al.}, we derive
an analytic expression of the MEF and the comparison of it with the
concurrence is given.

Quantum operation $\varepsilon^{Q}$ is a map for the state of $Q$
\begin{eqnarray}
\rho^{Q'}=\varepsilon^{Q}(\rho^{Q}).
\end{eqnarray}
Here $\rho^{Q}$ is the initial state of system $Q$, and after the
dynamical process the final state of the system becomes $\rho^{Q'}$.
Then the dynamical process is described by $\varepsilon^{Q}$. In the
most general case, the map $\varepsilon^{Q}$ must be a
trace-preserving and positive linear map \cite{stinespring,kraus},
so it includes all unitary evolutions. They also include unitary
evolving interactions with an environment $E$. Suppose that the
environment is initially in state $\rho^{E}$. The operator can be
written as
\begin{eqnarray}
\varepsilon^{Q}(\rho^{Q})&=&\tr_{E}{U(\rho^{Q}\otimes\rho^{E})U^{\dag}}\nonumber\\
&=&\tr_{E}{U(\rho^{Q}\otimes\sum_{i}{p_{i}|i\rangle\langle i|})U^{\dag}}\nonumber\\
&=&\sum_{j}{E_{j}^{Q}\rho^{Q} E_{j}^{Q\dag}}, \label{operation}
\end{eqnarray}
where $\sum_{i}{p_{i}|i\rangle\langle i|}$ is the spectral
decomposition of $\rho^{E}$, with $\{ |i\rangle \}$ being a base in
the Hilbert space $\mathcal{H}_{E}$ of the environment $E$, and
$E^{Q}_{j}=\sum_{i}{\sqrt{p_{i}}\langle j|U|i\rangle}$. Now we can
use Eq. ({\ref{operation}}) to get the intrinsic expression of
$\langle\Psi^{RQ}|\rho^{RQ'}|\Psi^{RQ}\rangle$, i.e.,
$F_{e}(\rho^{Q},\varepsilon^{Q})$. Because
\begin{eqnarray}
\rho^{RQ'}&= &\mathcal{I}^{R}\otimes\varepsilon^{Q}(\rho^{RQ})\nonumber\\
&=&\sum_{j}{(1^{R}\otimes E^{Q}_{j})\rho^{RQ}(1^{R}\otimes
E^{Q}_{j})^{\dag}}, \label{operation2}
\end{eqnarray}
one has
\begin{eqnarray}
F_{e}&=&\langle\Psi^{RQ}|\rho^{RQ'}|\Psi^{RQ}\rangle\nonumber\\
&=&\sum_{j}{\langle\Psi^{RQ}|(1^{R}\otimes
E^{Q}_{j})|\Psi^{RQ}\rangle}\nonumber\\
&~~~&~~~~~~{\times{\langle\Psi^{RQ}|(1^{R}\otimes
E^{Q}_{j})^{\dag}|\Psi^{RQ}\rangle}}\nonumber\\
&=&\sum_{j}{(\tr{\rho^{Q}E^{Q}_{j}})(\tr{\rho^{Q}E^{Q\dag}_{j}})}.
\label{the ef}
\end{eqnarray}
If systems $R$ and $Q$ both have zero internal Hamiltonian and there
is no interaction between $R$ and $Q$, the operation
$\varepsilon^{Q}$ entirely originates from the interaction between
$Q$ and the environment. In this sense the example of Jordan {\it et
al.} is a special case of this situation.

We consider two entangled qubits, $A$ and $B$, and suppose that
qubit $A$ interacts with a control qubit $C$. Then $A$, $B$ and $C$
respectively correspond to systems $Q$, $R$ and environment $E$ that
we have just referred. We suppose that the initial states of the
three qubits are
\begin{eqnarray}
W=\rho^{AB}_{\pm}\otimes\frac{1}{2} 1_{c} \label{w},
\end{eqnarray}
where
\begin{eqnarray}
\rho^{AB}_{\pm}=\frac{1}{4}(1\pm\sigma^{A}_{1}
\sigma^{B}_{1}\pm\sigma^{A}_{2}\sigma^{B}_{2}-\sigma^{A}_{3}\sigma^{B}_{3}),
\end{eqnarray}
with $\sigma^{A(B)}_{i}$, $i=1,2,3$, being Pauli matrices for qubit
$A(B)$. $\rho^{AB}_{+}$ and $\rho^{AB}_{-}$ are two Bell states,
representing the maximally entangled pure states for the combined
system of qubits $A$ and $B$. The total spins of states
$\rho^{AB}_{-}$ and $\rho^{AB}_{+}$ are 0 and 1, respectively.

We suggest an interaction between qubit $A$ and $C$ described by the
unitary transformation
\begin{eqnarray}
U=e^{-i t H} \label{u},
\end{eqnarray}
where
\begin{eqnarray}
H=\frac{\lambda\sigma^{A}_{3}}{2}(\ket{\alpha}\bra{\alpha}-\ket{\beta}\bra{\beta}),
\label{h}
\end{eqnarray}
$\lambda$ is the strength of the interaction, and $\ket{\alpha}$ and
$\ket{\beta}$ are two orthonormal vectors for system $C$. Then the
changing density matrix for the combined system of qubits $A$ and
$B$ can be calculated as
\begin{eqnarray}
\rho^{AB'}_{\pm}&=&\tr_{c}{\left[(U\otimes1^{B})W(U\otimes1^{B})^{\dag}\right]}\nonumber\\
&=&\frac{1}{4}[1\pm(\sigma^{A}_{1}
\sigma^{B}_{1}+\sigma^{A}_{2}\sigma^{B}_{2})\cos{(\lambda
t)}-\sigma^{A}_{3}\sigma^{B}_{3}]\nonumber\\
&=&\rho^{AB}_{\pm}\cos^{2}{(\frac{\lambda
t}{2})}+\rho^{AB}_{\mp}\sin^{2}{(\frac{\lambda t}{2})}. \label{dm}
\end{eqnarray}

The changing density matrix $\rho^{AB'}_{\pm}$ usually represents a
mixed state. In order to quantify the entanglement of it we use the
Wootters concurrence \cite{wootters1} defined as
\begin{eqnarray}
C(\rho)\equiv\max[0,\sqrt{\lambda_{1}}-\sqrt{\lambda_{2}}-
\sqrt{\lambda_{3}}-\sqrt{\lambda_{4}}],
\end{eqnarray}
where $\rho$ is the density matrix representing the investigated
state of the combined system of $A$ and $B$, $\lambda_{1}$,
$\lambda_{2}$, $\lambda_{3}$, and $\lambda_{4}$ are the eigenvalues
of
$\rho\sigma^{A}_{2}\sigma^{B}_{2}\rho^{\ast}\sigma^{A}_{2}\sigma^{B}_{2}$
in the decreasing order, and $\rho^{\ast}$ is the complex
conjugation of $\rho$. From Eq. (\ref{dm}) we can obtain
\begin{eqnarray}
C(\rho^{AB'}_{\pm})=|\cos{\lambda t}|.
\end{eqnarray}
It is found that at time $\lambda t=\frac{\pi}{2}$, the state
$\rho^{AB'}_{\pm}$ is changed from a maximally entangled state at
$t=0$ to a separable state and at time $\lambda t=\pi$ the state
$\rho^{AB'}_{\pm}$ returns to the maximally entangled state. The
explicit calculation about $\rho^{AB'}$ and $C(\rho^{AB'}_{\pm})$
can be seen in \cite{jordan}.

Now we adopt the EF to investigate this example. Using Eqs.
(\ref{operation}), (\ref{w}), (\ref{u}), and (\ref{h}), we obtain
the quantum operation on qubit $A$,
\begin{eqnarray}
\varepsilon^{A}(\rho^{A})&=&\tr_{C}{U(\rho^{A}\otimes\rho^{C})U^{\dag}}\nonumber\\
&=&\tr_{C}{U\left(\rho^{A}\otimes(\frac{1}{2}(\proj{\alpha}+\proj{\beta}))\right)U^{\dag}}\nonumber\\
&=&\frac{1}{2}e^{-i \sigma^{A}_{3}\left(\frac{\lambda
t}{2}\right)}\rho^{A}e^{+i \sigma^{A}_{3}\left(\frac{\lambda
t}{2}\right)}\nonumber\\
&~~~~&+\frac{1}{2}e^{+i \sigma^{A}_{3}\left(\frac{\lambda
t}{2}\right)}\rho^{A}e^{-i \sigma^{A}_{3}\left(\frac{\lambda
t}{2}\right)}. \label{expression}
\end{eqnarray}
So $E^{A}_{\alpha}=\frac{1}{\sqrt{2}}e^{-i
\sigma^{A}_{3}\left(\frac{\lambda t}{2}\right)}$ and
$E^{A}_{\beta}=\frac{1}{\sqrt{2}}e^{+i
\sigma^{A}_{3}\left(\frac{\lambda t}{2}\right)}$. Substituting them
into Eq. (\ref{the ef}) and noting that $\rho^{A} \equiv
\tr_{B}{(\rho^{AB}_{\pm})}=\frac{1}{2} 1$, we can get the EF as
\begin{eqnarray}
F_{e}&=&\sum_{j}{(\tr{\rho^{A}E^{A}_{j}})(\tr{\rho^{A}E^{A\dag}_{j}})}\nonumber\\
&=&\left(\frac{1}{\sqrt{2}}\tr\left[\left(
\begin{array}{c}
e^{-i\frac{\lambda t}{2}}~~~~~~0\\
~0~~~~~~~e^{+i\frac{\lambda t}{2}}
\end{array}\right)\frac{1}{2}\left(
\begin{array}{c}
1~~~~~~0\\
0~~~~~~1
\end{array}\right)\right]\right)^{2}\nonumber\\
&~~~~&+\left(\frac{1}{\sqrt{2}}\tr\left[\left(
\begin{array}{c}
e^{+i\frac{\lambda t}{2}}~~~~~~0\\
~0~~~~~~~e^{-i\frac{\lambda t}{2}}
\end{array}\right)\frac{1}{2}\left(
\begin{array}{c}
1~~~~~~0\\
0~~~~~~1
\end{array}\right)\right]\right)^{2}\nonumber\\
&=&\left(\cos\frac{\lambda t}{2}\right)^{2}. \label{solution}
\end{eqnarray}

\begin{figure}[h]
\includegraphics[width=0.8\columnwidth,
height=0.6\columnwidth]{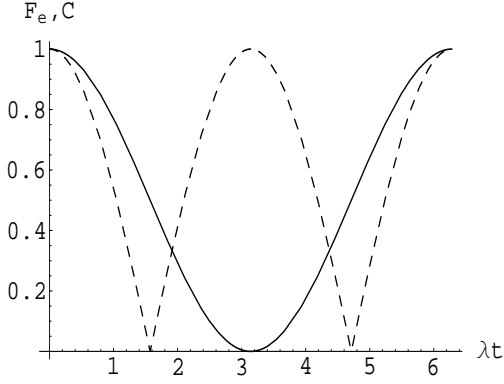} \caption{The evolutions of the EF
$F_{e}$ (solid line) and the concurrence $C$ (dashed line). We take
$\hbar=1$ so $\lambda t$ is dimensionless. } \label{fig}
\end{figure}

We can easily find the disagreement between the evolutions of
$F_{e}$ and $C(\rho^{AB'}_{\pm})$. At $\lambda t=\pi$, state
$\rho^{AB'}_{\pm}$ returns to the maximally entangled state as can
be seen from the concurrence, but its entanglement fidelity is zero
($F_{e}=0$). On the contrary, the initial maximally entangled state
have been changed to a separable state at $\lambda t=\frac{\pi}{2}$,
but the EF at this time is not zero. The evolutions of EF $F_{e}$
and concurrence $C(\rho^{AB'}_{\pm})$ are depicted in Fig.
\ref{fig}.

In fact,
$F_{e}(\rho^{Q},\varepsilon^{Q})=F_{s}^{2}(\rho^{RQ},\rho^{RQ'})$,
where $F_{s}(\rho^{RQ},\rho^{RQ'})$ is the static fidelity
\cite{nielsen}. The static fidelity satisfies $0\leq
F_{s}(\rho^{RQ},\rho^{RQ'})\leq 1$, where the first symbol of
``$\leq$" becomes equality if and only if $\rho^{RQ}$ and
$\rho^{RQ'}$ have orthogonal support, and the second symbol becomes
equality if and only if $\rho^{RQ}=\rho^{RQ'}$. When $\lambda
t=\pi$, from Eq. (\ref{dm}) we can see that
$\rho^{AB'}_{\pm}=\rho^{AB}_{\mp}$. The $\rho^{AB}_{\pm}$ are two
different Bell states and correspond respectively to eigenstates of
total spin one and total spin zero of the combined system of qubits
$A$ and $B$. So they have orthogonal support in the Hilbert space
$\mathcal {H}^{A}\otimes\mathcal {H}^{B}$. This is the reason for
the fact that
$F_{e}(\rho^{A},\varepsilon^{Q})=F_{s}^{2}(\rho^{AB},\rho^{AB'})=0$
at $\lambda t=\pi$.

The concept of the EF arises from the mathematical description for
the purification of mixed states. Any mixed state can be represented
as a subsystem of a pure state in a larger Hilbert space. The
entanglement of a pure state may cause the states of subsystems to
be mixed. The EF is usually used to measure how faithfully a channel
maintains the purification, or, equivalently, how well the channel
preserves the entanglement. In the above simple example, however, we
have found that, except for some special cases, only in the case of
slight noise, i.e., $\lambda t\longrightarrow0$, the EF
approximately agrees with the concurrence. This means that this
quantity may not be a good measurement for the evolution of the
entanglement in the processes of interaction with the environment.

In fact, Schumacher \cite{schumacher} has noted that the EF can be
lowered by a local unitary operation but the entanglement cannot be
so. From this consideration he defined the MEF
\begin{eqnarray}
F^{'}_{e}=\max_{U^{Q}}\bra{\Psi^{RQ}}(1^{R}\otimes
U^{Q})\rho^{RQ'}(1^{R}\otimes U^{Q})^{\dag}\ket{\Psi^{RQ}},
\label{f}
\end{eqnarray}
where $U^{Q}$ is any unitary transformation acting on $Q$. It is
clear that $F^{'}_{e}\geq F_{e}$. Since by using a proper local
unitary operation we can make the Bell state $\rho^{AB}_{\pm}$
become the Bell state $\rho^{AB}_{\mp}$, we can find that in the
above example $F^{'}_{e}=1$ at time $\lambda t=\pi$ whereas
$F_{e}=0$ at this time. So at $\lambda t=\pi$, the MEF equals the
concurrence. By using the quantum operation which we discussed
above, we can get the intrinsic expression of the MEF
\begin{eqnarray}
F^{'}_{e}&=&\max_{U^{Q}}\sum_{j}\bra{\Psi^{RQ}}(1^{R}\otimes
U^{Q}E_{j}^{Q})\ket{\Psi^{RQ}}\nonumber\\
&\times&\bra{\Psi^{RQ}}(1^{R}\otimes
U^{Q}E_{j}^{Q})^{\dag}\ket{\Psi^{RQ}}\nonumber\\
&=&\max_{U^{Q}}\sum_{j}(\tr{\rho^{Q}U^{Q}E_{j}^{Q}})(\tr{\rho^{Q}(U^{Q}E_{j}^{Q})^{\dag}}).
\end{eqnarray}

For this example we can derive an analytic expression of
$F^{'}_{e}$. Suppose $U$ is an arbitrary unitary operation on a
single qubit. Then it can be written as \cite{nielsen}
\begin{eqnarray}
U&=&e^{-i \alpha}R_{z}(\beta)R_{y}(\gamma)R_{z}(\delta)\nonumber\\
\nonumber\\
&=&e^{-i \alpha}\left(
\begin{array}{c}
e^{i (-\beta/2-\delta/2)}\cos\frac{\gamma}{2}~~~~~-e^{i (-\beta/2+\delta/2)}\sin\frac{\gamma}{2}\\
\\
e^{i (+\beta/2-\delta/2)}\sin\frac{\gamma}{2}~~~~~e^{i
(+\beta/2+\delta/2)}\cos\frac{\gamma}{2}
\end{array}\right), \nonumber
\end{eqnarray}
where $\alpha, \beta, \gamma$ and $\delta$ are real numbers, and
$R_{y(z)}$ is the rotation operator about the $y(z)$ axis. We have
\begin{eqnarray}
&\sum_{j}&(\tr{\rho^{A}U E_{j}^{A}})(\tr{\rho^{A}(U E_{j}^{A})^{\dag}})\nonumber\\
&=&\frac{1}{2}\left(\frac{1}{2}\tr \left(\begin{array}{c}
e^{i(-\beta/2-\delta/2-\frac{\lambda t}{2})}\cos\frac{\gamma}{2}~~~~~~0\\
\\
~0~~~~~~~e^{i(\beta/2+\delta/2+\frac{\lambda
t}{2})}\cos\frac{\gamma}{2}
\end{array}\right)\right)^{2}\nonumber\\
&~~~~&+\frac{1}{2}\left(\frac{1}{2}\tr \left(\begin{array}{c}
e^{i(-\beta/2-\delta/2+\frac{\lambda t}{2})}\cos\frac{\gamma}{2}~~~~~~0\\
\\
~0~~~~~~~e^{i(\beta/2+\delta/2-\frac{\lambda
t}{2})}\cos\frac{\gamma}{2}
\end{array}\right)\right)^{2}\nonumber\\
&=&\frac{1}{2}\cos^{2}(\frac{\gamma}{2})\cos^{2}(\beta/2+\delta/2+\lambda
t/2)\nonumber\\
&~~~~&+\frac{1}{2}\cos^{2}(\frac{\gamma}{2})\cos^{2}(\beta/2+\delta/2-\lambda
t/2).
\end{eqnarray}
\begin{figure}[h]
\includegraphics[width=0.8\columnwidth,
height=0.6\columnwidth]{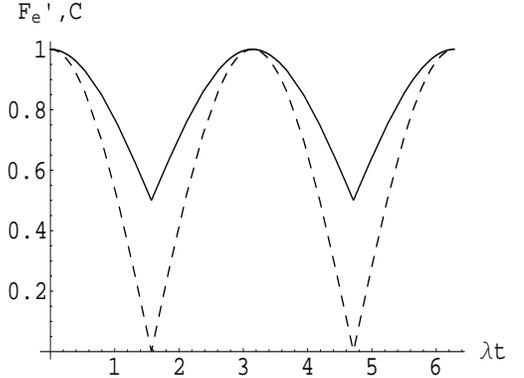} \caption{The evolutions of the
modified entanglement fidelity $F^{'}_{e}$ (solid line) and the
concurrence $C$ (dashed line).} \label{fig2}
\end{figure}
We should find a unitary operator $U$ which make
$\sum_{j}(\tr{\rho^{A}U E_{j}^{A}})(\tr{\rho^{A}(U
E_{j}^{A})^{\dag}})$ take its maximum value. Since
$\cos^{2}(\beta/2+\delta/2+\lambda t/2)\geq0$ and
$\cos^{2}(\beta/2+\delta/2-\lambda t/2)\geq0$, we can take
$\gamma=0$. So one obtains
\begin{eqnarray}
&\sum_{j}&(\tr{\rho^{A}U E_{j}^{A}})(\tr{\rho^{A}(U E_{j}^{A})^{\dag}})\nonumber\\
&=&1+\cos^{2}(\beta/2+\delta/2)(2\cos^{2}(\lambda
t/2)-1)\nonumber\\
&~~~~&-\cos^{2}(\lambda t/2).
\end{eqnarray}
When $2\cos^{2}(\lambda t/2)-1\geq0$ we take $
\cos^{2}(\beta/2+\delta/2)=1$ and get $F^{'}_{e}=\cos^{2}(\lambda
t/2)$; when $2\cos^{2}(\lambda t/2)-1 < 0$ we take
$\cos^{2}(\beta/2+\delta/2)=0$ and get $F^{'}_{e}=1-\cos^{2}(\lambda
t/2)$.

The evolutions of the MEF $F^{'}_{e}$ and the concurrence
$C(\rho^{AB'}_{\pm})$ are depicted in Fig. \ref{fig2}. We can find
that the MEF and the concurrence exhibit a similar behavior,
although their values do not exactly agree with each other at all
moments. When the state $\rho^{AB'}_{\pm}$ returns to the maximally
entangled state, the MEF is equal to $1$. The maximal difference
between them comes at the separable states where the MEF is equal to
$1/2$ while the concurrence is zero.

We have mentioned that the EF equals $1$ if and only if
$\rho^{RQ}=\rho^{RQ'}$. This means that the EF can be use to measure
the difference between a quantum channel and the identity channel.
If the concern is on the entanglement preserving in an evolution
process, however, one has to use the MEF because the EF can be
lowered by a local unitary operation in this process but the
entanglement cannot be so. If a quantum channel is just a unitary
operator, the entanglement is certainly invariant and the MEF always
equals to $1$ in the quantum process. In this sense the MEF can be
used to measure the difference between a quantum channel and an
arbitrary unitary operator.

In summary, for the example of Jordan {\it et al.}, we have derived
the analytic expressions of both the EF and the MEF, and show the
comparisons of them with the concurrence. From these we find that
the MEF may admirably reflects the entanglement preserving in a
quantum process.


\vskip 0.5 cm

{\it Acknowledgments} We wish to thank K. Surmacz for his
stimulating discussion which leads us to note the MEF. This work was
supported by National Foundation of Natural Science in China Grant
Nos. 60676056 and 10474033, and by the China State Key Projects of
Basic Research (2005CB623605 and 2006CB0L1000).


\end{document}